\documentclass{article}
\usepackage{spconf,amsmath,graphicx}
\usepackage{cite}

\title{NEW RESULTS ON MULTI-AGENT SYSTEM CONSENSUS: A GRAPH SIGNAL PROCESSING PERSPECTIVE}
\name{Jing-Wen Yi and Li Chai*\thanks{*Corresponding author. Thanks to the National Science Foundation of China (grant 61471275 and 61501337) and Hubei Provincial Education Department (grant T201302) for funding.}}
\address{School of Information Science and Engineering\\
Wuhan University of Science and Technology, Wuhan, China\\
\{yijingwen, chaili\}@wust.edu.cn}
\begin{document}
%
\maketitle
\begin{abstract}
This paper revisits the problem of multi-agent consensus from a graph signal processing perspective. By defining the graph filter from the consensus protocol, we establish the direct relation between average consensus of multi-agent systems and filtering of graph signals. This relation not only provides new insights of the average consensus, it also turns out to be a powerful tool to design effective consensus protocols for uncertain networks, which is difficult to deal with by existing time-domain methods. In this paper, we consider two cases, one is uncertain networks modeled by an estimated Laplacian matrix and a fixed eigenvalue bound, the other is connected graphs with unknown topology. The consensus protocols are designed for both cases based on the protocol filter. Several numerical examples are given to demonstrate the effectiveness of our methods.
\end{abstract}
\section{Introduction}

Consensus of multi-agent systems (MASs) is a fundamental problem in collective behaviors of autonomous individuals, which has been extensively studied in the last decade  \cite{[01]Olfati, [02]YouLiXie, [3]ZhuHongChen}. The key problem for consensus is to design appropriate distributed protocols that each agent can only get information from its local neighbors, and the whole network of agents may coordinate to reach an agreement on certain quantities of interest eventually. Most of previous researches analyzed and solved the consensus problem based on the time-domain state-space model \cite{[03]LiZhang, [04]LiuXie, [05]Seyboth}. Other studies utilized the Nyquist stability criterion in the frequency domain \cite{[06]OlfatiShamma, [07]TianLiu}. One contribution of this paper is to provide new insights of the consensus problem by exploring the recent development of graph signal processing in spatial frequency domain.

Graph signal processing has drawn great interests for analyzing high-dimensional data in recent years \cite{[11]ShumanOrtega, [08]Moura, [9]SakiyamaTanaka}. There are two different frameworks on graph signal processing. One develops the discrete signal processing structure and concepts on graph based on the Jordan normal form and generalized eigenbasis of the adjacency matrix\cite{[08]Moura, [12]Moura, [15]Moura, [16]Moura, [17]Moura}. The other establishes the framework by merging algebraic and spectral graph concepts with classical signal processing which interprets the graph Laplacian eigenvalues as graph frequencies and the orthonormal eigenvectors as graph Fourier transformation \cite{[11]ShumanOrtega, [13]Ortega, [18]Ortega, [19]Ortega, [20]Ortega}.

In this paper, we will first reveal the explicit connection between filtering of graph signals and consensus of MASs. It is shown that the MAS can reach its average consensus in finite time by designing an appropriate protocol filter to keep only the low frequency component of the graph (corresponding to the zero eigenvalue of the graph Laplacian matrix) while surrogating other higher frequency components. The designed graph filter can be implemented by a distributed consensus protocol derived from the closed-loop property of the MAS and the Laplacian matrix of the network graph.

Viewing MASs from graph signal processing perspective not only provides new insights, it also presents a new methodology to solve some challenging problems in MASs on uncertain networks. For MASs with estimated Laplacian matrix, we show that the graph signal processing perspective can help to design distributed consensus protocol gains as well as estimate the asymptotic consensus error, which is difficult to analyze by the existing time-domain state-space model based methods. For MASs with completely unknown Laplacian matrix except assuming the connectivity and the maximum degree of the network graph, we provide a new design method of the consensus protocol gain. The consensus error bound is also presented. Numerical examples are given to demonstrate the effectiveness of the proposed methods.

\section{PRELIMINARIES}

\subsection{Spectral Graph Theory}
Let $\mathcal{G} = (\mathcal{V},\mathcal{E} ,\mathcal{A})$ be an undirected graph of order $N$ ($N \ge 2$) which consists of an agent set $\mathcal{V} = \{ {{\rm{v}}_1},{{\rm{v}}_2}, \ldots ,{{\rm{v}}_N}\} $, an edge set $\mathcal{E} \subseteq \mathcal{V} \times \mathcal{V}$, and a adjacency matrix $\mathcal{A} = [{a_{ij}}] \in {\Re ^{N \times N}}$. An edge ${e_{ij}} = ({{\rm{v}}_i},{{\rm{v}}_j}) \in \mathcal{E}$ if and only if $a_{ij}=a_{ji} \ne 0$, that means there exist communications between agent $i$ and agent $j$. Moreover, self-edges are not allowed, i.e.,  ${e_{ii}} = ({{\rm{v}}_i},{{\rm{v}}_i}) \notin {\cal E}$ and ${a_{ii}} = 0$. The Laplacian matrix of graph is defined by $\mathcal{L} = \mathcal{D} - \mathcal{A}$, where $\mathcal{D} \buildrel \Delta \over = diag\{ d_1, \ldots ,d_N\}$ is the degree matrix, and $d_i=\sum\nolimits_{j = 1}^N {{a_{ij}}}$. The Laplacian matrix $\mathcal {L}$ is symmetric and all the eigenvalues are real.

\textbf{Lemma 1.} \cite{[14]Bullo} For an undirected graph $\mathcal{G}$, the Laplacian matrix $\mathcal{L}$ has at least one zero eigenvalue, and all the non-zero eigenvalues are positive, i.e., $0={\lambda _{1}} \le {\lambda _{2}} \le \cdots  \le {\lambda _{N}}$. Furthermore, zero is a simple eigenvalue of $\mathcal{L}$ with the associated orthonormal eigenvector $\frac{1}{{\sqrt N }}{{\vec 1} _N}$, where ${{\vec 1}_N}\buildrel \Delta \over =[1,\ldots,1]^T \in \Re ^ N$, if and only if the $\mathcal{G}$ is connected.

\subsection{Graph Signal Processing}
Consider a graph $\mathcal{G} = (\mathcal{V},\mathcal{E} ,\mathcal{A})$, its Laplacian matrix can be written as $\mathcal{L}=V \Lambda V^T$, where $\Lambda=diag\{\lambda_1,\lambda_2,\ldots,\lambda_N\}$, $V=[{\vec v}_1,{\vec v}_2,\ldots,{\vec v}_N] \in \Re ^ {N \times N}$ is a unitary matrix. The graph signal $x$ is the collection of the signal values on all the agents, i.e., $x=[x_1,x_2,\cdots,x_N]^T \in \Re ^ N$. Similarly to classical signal processing, the Fourier transform in the graph signal processing \cite{[11]ShumanOrtega} is defined on the graph spectra as $\hat {x}={V^T}x$, and the inverse graph Fourier transform is given by $x={V}{\hat {x}}$, where $\hat {x}=[\hat{x}_1, \hat{x}_2, \cdots, \hat{x}_N]^T \in \Re ^ N$.

Let $h(\cdot)$ be the transfer function of a filter, and $\hat y =[{\hat y}_1,{\hat y}_2,\ldots,{\hat y}_N]^T \in \Re ^ N$ be the filtered graph signal in spatial frequency domain, then the graph spectral filtering can be defined as $\hat {y}_i = h(\lambda _i) \hat {x}_i$.
Taking the inverse graph Fourier transform, the filtered graph signal in time domain can be obtained as $y = V diag\{{h({\lambda _1})}, \cdots, {h({\lambda _N})}\}{V^T}x$, where $y =[{y}_1,{y}_2,\ldots,{y}_N]^T \in \Re ^ N$ is the filted graph signal in time domain. 


\section{MULTI-AGENT CONSENSUS VIA GRAPH SIGNAL PROCESSING}

Consider the dynamics of the MAS on a graph $\mathcal{G} = (\mathcal{V},\mathcal{E} ,\mathcal{A})$ with $N$ agents described by
\begin{equation}\label{eq6}
{x_i}(k + 1) = {x_i}(k) + {u_i}(k), \quad i=1,2,\ldots,N
\end{equation}
where ${x_i}(k)$, ${u_i}(k) \in \Re$ is the state and the control input, respectively. A commonly used control protocol \cite{[01]Olfati} is as follows 
\begin{equation}\label{eq7}
{u_i}(k) = {\varepsilon _k} \sum\limits_{j \in {N_i}} {{a_{ij}}({x_j}(k) - {x_i}(k))},
\end{equation}
where ${\varepsilon _k}$ is the control gain at time $k$, and $\mathcal{A}=[a_{ij}]_{N \times N}$ is the adjacency matrix of graph $\mathcal{G}$.

\textbf{Definition 1.} For a control protocol  within time $T$ given by (\ref{eq7}), the corresponding protocol filter is defined as
\begin{equation}\label{eq07}
h(\lambda, T)={\prod\limits_{k = 0}^{T-1} {(1 - {\varepsilon _k}{\lambda})} }.
\end{equation}

The average consensus of multi-agent system (\ref{eq6}) under protocol (\ref{eq7}) is said to be achieved asymptotically if $\mathop {\lim }\limits_{k \to \infty } {x_i}(k) = \frac{1}{N}\sum\limits_{i = 1}^N {{x_i}(0)}$, $i=1,2,\ldots,N$. And the average consensus is said to be reached at time $T$ if ${x_i}(k) = \frac{1}{N}\sum\limits_{i = 1}^N {{x_i}(0)}$, $ i=1,2,\ldots,n$, hold for all $k \ge \rm T$.

From (\ref{eq6}) and (\ref{eq7}), the dynamics of the multi-agent system can be calculated that
\begin{equation}\label{eq8}
\begin{array}{l}
x(k + 1) = (I - {\varepsilon _k}L)x(k)\\
 = V diag\{h({\lambda_1},k), \cdots, h({\lambda_N},k)\}{V^T}x(0).
\end{array}
\end{equation}
Let the initial state $x(0)$ and the current state $x({k+1})$ of the MAS be the collective signal values of the original graph signal and the filtered one, respectively. Then, the MAS plays a role of the filter for the graph signal with $x(0)$ and $G$, and the the transfer function of the filter is $h({\lambda },k)$ defined in Definition 1. The following result gives the property of the protocol filter achieving average consensus.

\textbf{Theorem 1.} For the MAS (\ref{eq6}) on a connected graph $\mathcal{G}$, of which the Laplacian matrix has eigenvalues as $0=\lambda_1<\lambda_2 \le \ldots \le \lambda_N$, assume the consensus protocol is in the form of (\ref{eq7}). Then the MAS reaches average consensus at time $T$ if and only if the corresponding protocol filter $h(\lambda,T)$ defined by (\ref{eq07}) satisfies $h(0, T)=1$ and $h({\lambda_i}, T)=0$ for $i = 2, \ldots ,N$.

The proof of Theorem 1 is omitted due to space limitation. It is easy to verify that the consensus state is $x(T)={{\vec v}_1}{\vec v}_1^Tx(0)$. Theorem 1 shows that the protocol filter can be viewed as a low-pass filter with zero at high frequency components $\lambda_2,\ldots,\lambda_N$. It follows from Theorem 1 that the MAS with $N$ agents on a connected graph can definitely reach the average consensus at time $N-1$ by properly choosing the control gain $\varepsilon_k$. The following corollary shows that the consensus time can be smaller than $N-1$.

%

\textbf{Corollary 1.} For the MAS (\ref{eq6}) on a connected graph with $p$ distinct nonzero eigenvalues ($0=\lambda_1<\lambda_2<\cdots<\lambda_{p+1}$), take the control gains as
\begin{equation}\label{eq12}
{\varepsilon_k} = \left\{ {\begin{array}{*{20}{c}}
{\frac{1}{{{\lambda _{p+1-k}}}},\quad \quad k = 0,1, \ldots , p-1},\\
{0,\quad \quad \quad \quad \quad {\rm{otherwise}}.\quad }
\end{array}} \right.
\end{equation}
Then the consensus protocol (\ref{eq7}) makes the MAS reach average consensus at time $p$, that is, $x(p)=\frac{1}{N}\sum\limits_{i = 1}^N {{x_i}(0)}$.

\textbf{Remark.} The fact that finite time consensus can be achieved by choosing the control gains equal to the reciprocal of nonzero Laplacian eigenvalues is not new. It has been obtained by using different methods, for example, matrix factorization method \cite{[21]Hendrickx}, minimal polynomial method \cite{[22]Kibangou}. By defining the protocol filter $h(\lambda,T)$, Theorem 1 derives the consensus result from a graph signal processing perspective. In the next section, we will show that this method is a powerful tool to solve the consensus of MASs on uncertain networks, which is difficult (sometimes unable) to deal with by existing methods.

%

\section{CONSENSUS ON UNCERTAIN NETWORKS}

In the previous section, we assume that the graph structure of MASs is completely known. The exact average consensus in finite-time can be reached by designing a appropriate protocol filter. This section will discuss the average consensus problem in two non-ideal cases: estimated Laplacian matrix and unknown network topology. Denote the consensus error as $e(t) = \sum\limits_{i = 1}^N {{{({x_i}(t) - \frac{1}{N}\sum\limits_{i = 1}^N {{x_i}(0)})^2}}}$, the consensus performance will be analyzed in both cases.


\subsection{Consensus for MASs on Graphs with Estimated Laplacian Matrix}

Consider the MAS on a connected graph $\tilde {\mathcal{G}}$ with Laplacian matrix $\tilde {\mathcal{L}}$. Denote the eigenvalues of $\tilde {\mathcal{L}}$ as $0={\tilde {\lambda} _1}<{\tilde {\lambda} _2} \le \ldots \le {\tilde {\lambda} _N}$. Let $\mathcal{L}_0=V {\Lambda_0} {V^T}$ be the estimated Laplacian matrix of $\tilde {\mathcal{L}}$, where ${\Lambda_0} = diag\{{\lambda _1},{\lambda _2}, \ldots ,{\lambda _N}\}$. For $\bar \delta >0$, denote
\begin{equation}\label{eq11}
\varphi = \frac{{\bar \delta}^2}{{{{\lambda }_2}^2}} \prod\nolimits_{k = 1}^{N-2} (1 - \frac{{{\lambda}_2}+\bar \delta}{{{{\lambda }_{k+2}}}})^2.
\end{equation}


%


\textbf{Theorem 2.} For the MAS (\ref{eq6}) on the graph $\tilde {\mathcal{G}}$ under the consensus protocol (\ref{eq7}), take the control gain as
\begin{equation}\label{eq13}
{\varepsilon _{k + j(N-1)}} = \begin{array}{*{20}{c}}
{\frac{1}{{{\lambda _{N - k}}}},}&\begin{array}{c}
k = 0,1, \ldots ,N - 2,\\
j = 0,1, \ldots ,\infty ,
\end{array}
\end{array}
\end{equation}
where ${\lambda _2}, \ldots ,{\lambda _N}$ are eigenvalues of the estimated Laplacian matrix $\mathcal{L}_0$. Then the MAS reaches average consensus asymptotically if $\left| {{{\tilde \lambda }_i} - {\lambda _i}} \right| \le \bar \delta$ and $\varphi <1$, where $\varphi$ is defined in (\ref{eq11}). Moreover, at time $T_j=j(N-1)$, the consensus error satisfies $e(T_j) \le {\varphi^j} {\left\| {x(0)} \right\|^2}$.

\emph{Proof Outline:} The corresponding protocol filter at each control period can be written as $h(\lambda, j{(N-1)})=\prod\limits_{k = 0}^{N-2} (1 - \frac{\lambda}{{{\lambda _{N - k}}}})^j $. Then the consensus error at time $T_j=j(N-1)$ can be calculated by $e(T_j) = \left\| \sum\limits_{i = 2}^N {h({\lambda _i}, {T_j}){{\vec v}_i}{\vec v}_i^T} x(0) \right\|^2 \le {\varphi^j} {\left\| {x(0)} \right\|^2}$. It is easy to see that $\mathop {\lim }\limits_{j \to \infty } e({T_j}) = 0$ since $\varphi <1$. Thus, the MASs (\ref{eq6}) can reach the average consensus asymptotically by the protocol (\ref{eq7}).

\begin{figure}[htb]
  \centering
  \centerline{\includegraphics[width=8cm]{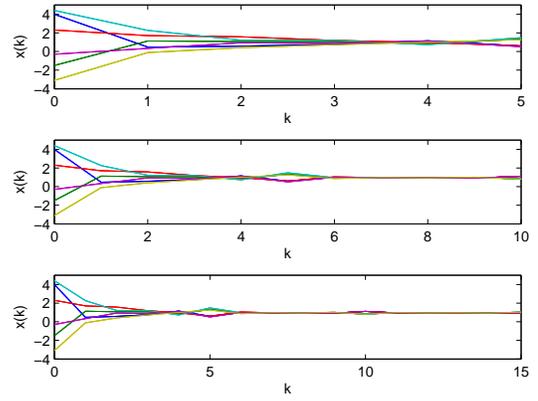}}
  \caption{(a) The trajectory of agent state $x(k)$ for $T_s=5$. (b) The trajectory of agent state $x(k)$ for $T_s=10$. (c) The trajectory of agent state $x(k)$ for $T_s=15$.}
  \label{fig2}
\end{figure}

\begin{figure}[!htb]
  \centering
  \centerline{\includegraphics[width=8cm]{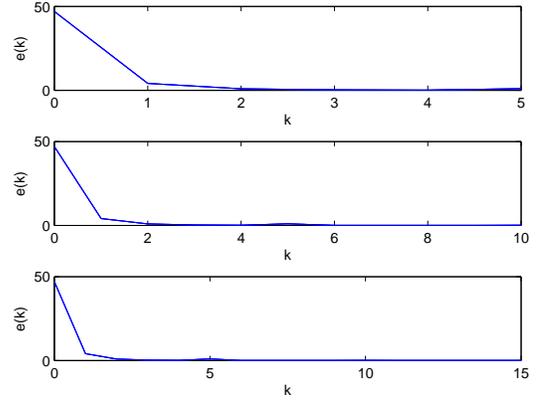}}
  \caption{(a) The consensus square error $e(k)$ for $T_s=5$. (b) The consensus square error $e(k)$ for $T_s=10$. (c) The consensus square error $e(k)$ for $T_s=15$.}
  \label{fig2}
\end{figure}

\emph{Example 1:} Consider the MAS on a connected graph $\tilde {\mathcal{G}}$ with $6$ agents, the Laplacian matrix is given as $\tilde {\mathcal{L}}=\mathcal{L}_0+\bar \delta V diag\{0,I_{N-1}\} V^T$, where $\mathcal {L}_0 = V {diag\{0, 1, 1, 3, 3, 4\}} V^T$ is the estimated Laplacian matrix corresponding to an unweighted cycle network, and $\bar \delta =0.5$ is the estimation error bound. From Theorem 2, the control gain in one control period can be designed as ${\varepsilon _{k}} =\frac{1}{4}, \frac{1}{3}, \frac{1}{3}, 1, 1$ for $k=0,1,2,3,4$. Let the simulation time be $T_s=5, 10, 15$, respectively, then the evolution of agent states and the consensus error are shown in Fig. 1 and Fig. 2. It can be seen that the MAS (\ref{eq6}) under the designed protocol (\ref{eq7}) reaches average consensus asymptotically, and the consensus errors at each $T_s$ are $e(5)=0.9675$, $e(10)=0.1376$, and $e(15)=0.0017$.

\subsection{Consensus for MASs with Unknown Network Topology}

Consider a connected graph $\mathcal {G}$ with unknown network topology, the methods proposed in previous sections have no application to solve the average consensus problem in this case. Assume the maximum degree of the graph is given, i.e., $\mathop {\max }\limits_{i = 1, \ldots ,N} \{ {d_i}\} = \bar d$. Then, divide $[0,2{\bar d}]$ into $T_0$ uniform intervals, and the length of each interval is $\frac{{2\bar d}}{T_0}$. A general result for average consensus can be derived as follows.


\textbf{Theorem 3.} For the MAS (\ref{eq6}) on an unknown connected graph $\mathcal {G}$ with maximum degree $\bar d$, take the control gains as
\begin{equation}\label{eq14}
{\varepsilon _{k + j{T_0}}} = \begin{array}{*{20}{c}}
{\frac{{T_0}}{{2({T_0} - k){\bar d} }},}&\begin{array}{c}
k = 0,1, \ldots ,{T_0} - 1,\\
j = 1, \ldots \infty .
\end{array}
\end{array}
\end{equation}
Then the consensus protocol (\ref{eq7}) makes the MAS reach average consensus asymptotically. Furthermore, assume the lower bound of the algebraic connectivity satisfies ${\lambda_2} \ge \frac{2 {\bar d}}{\alpha {T_0}}$, then the consensus error at time $T_j=j{T_0}$ satisfies $e(j{T_0})\le \varphi^{2j}{\left\| {x(0)} \right\|^2}$, where $\varphi=\max \{ 1 - \frac{1}{\alpha}\ln ({T_0} + 1),\frac{1}{{2{T_0}}}\} <1$.

\emph{Proof Outline:} From the design of the control gain, the corresponding protocol filter at the end of each control period can be derived as $h(\lambda, j{T_0})={\prod\limits_{k = 0}^{{T_0}-1} {(1 - \frac{\lambda{T_0}}{2({T_0}-k)\bar d})^j} }$. It can be calculated that $\left| {h({\lambda}, T_0)} \right| \le \psi$, where $\psi = \mathop {\max }\limits_{\lambda  \in (0,2\bar d]} \{ \prod\limits_{k = 0}^{{T_0} - 1} (1 - \frac{\lambda }{{({T_0} - k)\phi }})\}  <1$. And the consensus error can be calculated as $e(j{T_0}) = {\left\| {\sum\limits_{i = 2}^N {h({\lambda _i}, j{T_0}){{\vec v}_i}{\vec v}_i^T} x(0)} \right\|^2}\le \psi^{2j}{\left\| {x(0)} \right\|^2}$.  It is easy to see that $\mathop {\lim }\limits_{j \to \infty } e(j{T_0}) = 0$ since $\psi <1$. Then the MAS reaches the average consensus asymptotically under the protocol (\ref{eq7}).

Assume the lower bound of the algebraic connectivity satisfies ${\lambda_2} \ge \frac{2 {\bar d}}{\alpha {T_0}}$, a more accurate upper bound of the protocol filter can be derived as $\left| {h({\lambda}, T_0)} \right| \le \max \{ 1 - \frac{1}{\alpha}\ln ({T_0} + 1),\frac{1}{{2{T_0}}}\}$. Then the consensus error at time $T_j=j{T_0}$ satisfies $e({T_j}) \le \varphi^{2j}{\left\| {x(0)} \right\|^2}$.

It follows from Theorem 3 that the algebraic connectivity of the graph plays an important role in reaching consensus, and the higher algebraic connectivity corresponding to a better consensus performance and a lower consensus error at each control period.

\emph{Example 2:} Consider a MAS with $6$ agents on two different graphs, one is an unweighted cycle $\mathcal{G}_1$, the other is an unweighted path $\mathcal{G}_2$. The maximum degrees of the two graphs are the same, i.e., ${\bar d}_1={\bar d}_2=2$. Divide $[0,4]$ into $5$ uniform intervals, then ${T_0}=5$ and $\frac{{2\bar d}}{T_0}=0.8$. From Theorem 3, the control gain in one period can be derived as $\varepsilon _{k}= \frac{1}{4-0.8k}$, $k=0,\ldots,4$. For the MAS (\ref{eq6}) on two graphs $\mathcal{G}_1$ and $\mathcal{G}_2$ respectively, the protocol (\ref{eq7}) with the designed periodic control gain can solve the average consensus asymptotically as shown in Fig 3. Moreover, it is easy to verify that the algebraic connectivity of the cycle is higher than that of the path, thus the consensus performance of the graph $\mathcal{G}_1$ is much better than the graph $\mathcal{G}_2$, and the consensus errors of the two graphs are shown in Fig 4.




\begin{figure}[htb]
  \centering
  \centerline{\includegraphics[width=8cm]{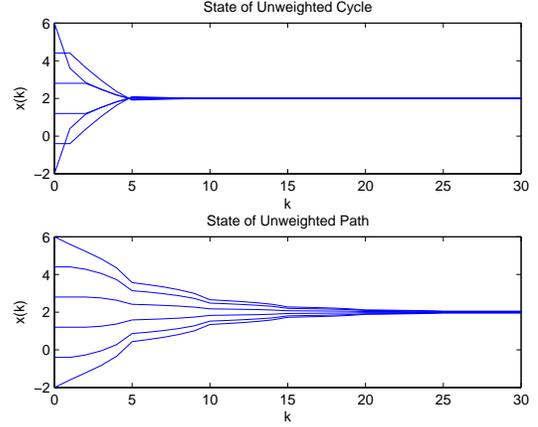}}
  \caption{(a) The agent state trajectory of the unweighted cycle $\mathcal{G}_1$. (b) The agent state trajectory of the unweighted path $\mathcal{G}_2$.}
  \label{fig4}
\end{figure}

\begin{figure}[!htb]
  \centering
  \centerline{\includegraphics[width=8cm]{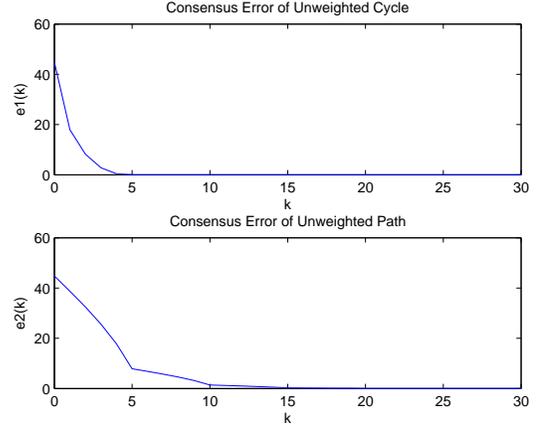}}
  \caption{(a) The consensus error $e_1(T)$ of the unweighted cycle $\mathcal{G}_1$. (b) The consensus error $e_2(T)$ of the unweighted path $\mathcal{G}_2$.}
  \label{fig5}
\end{figure}


\section{CONCLUSION}

This paper has established the explicit connection between filtering of graph signals and consensus of MASs. It has been shown that the MAS can reach its average consensus in finite time by designing an appropriate protocol filter, which can be implemented by a distributed consensus protocol. By using the concept of the protocol filter, we provide new methods to solve the average consensus problem in cases of estimated Laplacian matrix and unknown network topology. The asymptotic consensus error has been analyzed in both cases. While the protocol filter is defined for MASs on undirected graphs, it can be easily extended to direct graphs. We only consider MASs consisting of agents with first-order dynamics, it is interesting to extend our methods to MASs with high-order dynamics.

\bibliographystyle{unsrt}

\end{document}